\documentclass[preprint]{iucr}              
\usepackage{xcolor}
\usepackage{tabularx}
\usepackage{amssymb,amsmath}
\usepackage{gensymb}

     \journalcode{S}              

\begin{document}                  



\title{Artifact Identification in X-ray Diffraction Data using Machine Learning Methods}


\cauthor[a]{Howard}{Yanxon}{yanxonh@anl.gov}{address if different from \aff}
\author[a]{James}{Weng}
\author[a]{Hannah}{Parraga}
\author[a]{Wenqian}{Xu}
\author[a]{Uta}{Ruett}
\author[a]{Nicholas}{Schwarz}

\aff[a]{Argonne National Laboratory, 9700 South Cass Avenue, Lemont, IL 60439, \country{USA}}









\maketitle

\begin{abstract}
The in situ synchrotron high-energy X-ray powder diffraction (XRD) technique is highly utilized by researchers to analyze the crystallographic structures of materials in functional devices (e.g., battery materials) or in complex sample environments (e.g., diamond anvil cells or syntheses reactors). An atomic structure of a material can be identified by its diffraction pattern, along with detailed analysis such as Rietveld refinement which indicates how the measured structure deviates from the ideal structure (e.g., internal stresses or defects). For in situ experiments, a series of XRD images is usually collected on the same sample at different conditions (e.g., adiabatic conditions), yielding different states of matter, or simply collected continuously as a function of time to track the change of a sample over a chemical or physical process. In situ experiments are usually performed with area detectors, collecting 2D images composed of diffraction rings for ideal powders. Depending on the material's form, one may observe different characteristics other than the typical Debye Scherrer rings for a realistic sample and its environments, such as textures or preferred orientations and single crystal diffraction spots in the 2D XRD image. In this work, we present an investigation of machine learning methods for fast and reliable identification and separation of the single crystal diffraction spots in XRD images. The exclusion of artifacts during an XRD image integration process allows a precise analysis of the powder diffraction rings of interest. We observe that the gradient boosting method can consistently produce high accuracy results when it is trained with small subsets of highly diverse datasets. The method dramatically decreases the amount of time spent on identifying and separating single crystal spots in comparison to the conventional method.

\end{abstract}


\section{Introduction}

X-ray user facilities produce some of the largest scientific data in the world. The experiments performed at these facilities cover a large range of multi-disciplinary sciences and produce data which are often a sum of various materials in the path of the X-ray beam. The identification of the data of interest, assurance of sufficient data quality, and recognition of systematic errors requires a high level of expertise. It is therefore essential to coordinate data collection and analysis to achieve optimized efforts in scientific research. Many endeavors have been made to optimize data acquisition processes to improve scientific reproducibility, transparency, and reliability overcoming limitations in data quality to open paths to new discoveries. The success of these efforts will enable users to efficiently search and effectively comprehend the necessary measures for understanding the complex concepts of natural phenomena.
 
Artificial intelligence/machine learning (AI/ML) advances have shown promises, not only in speeding up, but also expanding the robustness of data analysis methods, and are poised to play an important role in X-ray user facilities \cite{sivaraman2021experimentally, sivaraman2021automated, du2020three, cherukara2018real, liu2022braggnn, yang2018low, liu2020tomogan}. For example, PtychoNN---a Python library that leverages deep convolutional neural networks (CNN) to solve image reconstruction problems in Ptychograpic X-ray imaging---has improved the speed by two orders-of-magnitude and with 5x less data than that required by current methods which use phase retrieval algorithms \cite{cherukara2020ai}. An approach that uses a ML clustering method to remove spurious data in Bragg coherent diffraction imaging has decreased the amount of time spent in manual data processing \cite{pelzer2021removal}. Additionally, Grazing Incidence Small Angle X-ray Scattering (GISAXS) is a surface-sensitive technique that has seen a tremendous growth in popularity in probing complex morphologies ranging from the fields of polymer and soft matter science to hard condensed matter \cite{hexemer2015advanced}. The scattering patterns can hold a variety of features, such as peaks, rods, and rings, whose position in its reciprocal space can often express its real space features. Several use cases of ML techniques have been successfully applied to other studies similar to the GISAXS type of problem \cite{wang2017x, kiapour2014materials, ziletti2018insightful}. Another example of ML application is illustrated in post data analysis of X-ray diffraction (XRD) experiments. ML methods have been developed and compared to classify crystal structure symmetries and space groups from simulated and experimental 1D XRD patterns, resulting in interpretable outcomes \cite{oviedo2019fast, suzuki2020symmetry}. Deep variational autoencoders applied to simulated and experimental 1D XRD patterns can be used for classification and XRD reconstructions in a Co-Cr-Ni-Re system composing in three phase mixtures \cite{banko2021deep}. Finally, phase maps of a Fe-Co-Ni ternary alloy can be constructed by measuring the dissimilarity among the 1D XRD patterns \cite{iwasaki2017comparison}.

Recent advances in ML techniques have had tremendous impact across industry and scientific domains, especially in the application of image processing. For instance, CNN and gradient boosting are successfully applied and used in image segmentation \cite{badrinarayanan2017segnet, chen2014semantic, yu2015multi} and image recognition \cite{simonyan2014very, he2016deep, peter2017cost}. These methods require diverse datasets on which the algorithms can train to select values through potentially millions of hyperparameters required to produce accurate models. The datasets are necessary to provide sufficient samplings of the relevant feature spaces. The ML recognition and segmentation applications can be carried over to other datasets such as XRD images.

In this paper, we will demonstrate the capability of ML methods for recognizing and segmenting artifacts that appear in a typical 2D XRD image. The in situ synchrotron high-energy XRD technique \cite{ren2012high} is widely used to study the crystallographic structures of inorganic materials. A crystallographic structure of a material is identified by its diffraction pattern, along with Rietveld analysis which indicates how the measured structure deviates from the ideal structure (e.g., internal stresses or defects). XRD images/patterns can be collected under different conditions such as pressure, temperature, electrical field, and/or gas atmosphere, which can yield different states of matter. Since the Advance Photon Source (APS) has a plethora of XRD images collected by many user groups, XRD experiments are well suited to benefit by exploiting ML methods.

2D XRD images are usually composed of many characteristics such as rings, grainy rings, preferred orientations or texture, and single crystal diffraction spots. These characteristics can be identified by their positions and intensities. For example, single crystal diffraction spots usually span a small region in an XRD image with very high intensity. In many situations, it is desirable to separate single crystal diffraction spots before the integration procedure in order to produce the conventional 1D XRD pattern with accurate peak intensities and profiles. Software such as GSAS--II \cite{toby2013gsas}---a state-of-the-art software for crystallographic structural analysis---has a segmentation algorithm that can automatically identify single crystal diffraction spots based on image pixel intensity. Nevertheless, the algorithm can fail to differentiate the artifacts when other characteristics, such as preferred orientations, are also present in an XRD image.

The performance of several ML methods is compared based on their abilities to mask out single crystal diffraction spots. The demonstration shows that ML methods are suitable approaches to recognize and separate (i.e., mask) in both speed and accuracy despite a number of confounding factors that may impact the accuracy of results, such as peak shifting due to lattice expansion or contraction. In addition, the speed-up makes ML methods applicable for on-the-fly masking during an XRD experiment. This implementation not only reduces the required person-power to analyze data, but also enables correct on-the-fly data analysis optimizing data collection.

\section{Data and Methods}

\subsection{Datasets}\label{datasets}

\begin{figure}
\centering
\caption{(a) A raw experimental XRD pattern and (b) its masking result. The red, blue and green boxes in (a) show (c) single crystal diffraction spots, (d) preferred orientation (the darker line), and two texture lines. (a) and (b) are cropped for visualization purposes.}
\includegraphics[width=0.95\textwidth]{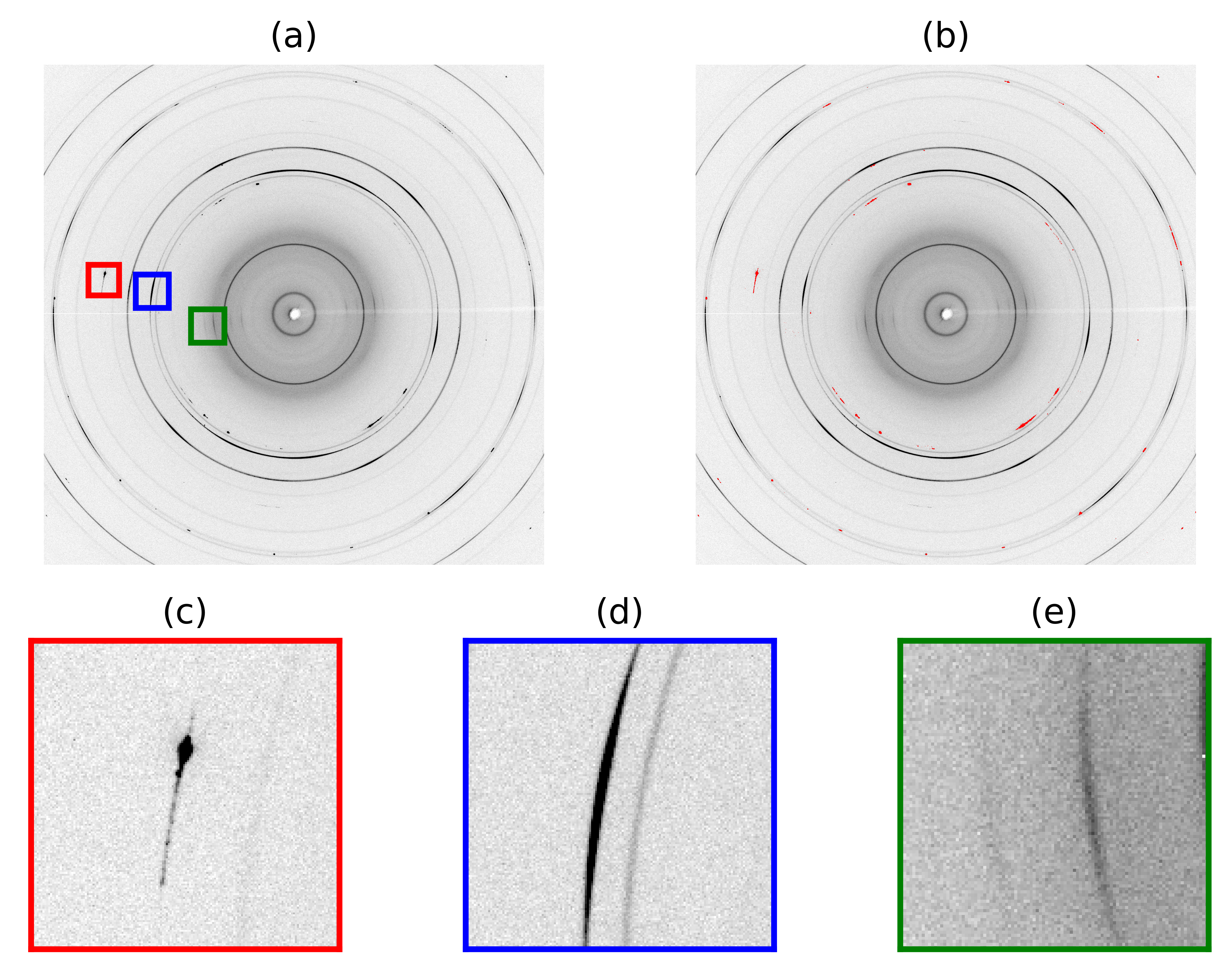}
\label{xrd_example}
\end{figure}

For this study, we use 5 diverse datasets that are comprised of 2D raw XRD images belonging to different material compositions and collected under different experimental conditions at the APS Beamline 17-BM (see Table \ref{datasets_table}). The raw images are 2880x2880 pixel intensity arrays, collected with a Varex XRD 4343CT area detector. As mentioned above, an XRD image can contain different characteristics as seen in Fig. \ref{xrd_example}. In the figure, the intensities of the preferred orientation and single crystal diffraction spots are often in the same order-of-magnitude. The preferred orientations show differential intensities around a diffraction ring and are usually symmetric. This means that an intense band at a particular azimuth angle is met with a similarly intense band in the opposite or 180\degree direction, whereas the single crystal diffraction spots can be due to scattering from the sample holder, single crystal phases formed in a reaction, etc. Each of the datasets is obtained under unrelated experimental conditions, such as different wavelength, detector center, distance of sample to detector, etc. The information is stored as metadata along with the detector image. Meanwhile, the overall scale of pixel intensities in an XRD image is related to the inlet photon flux of the beam, the amount of sample in the beam path, the sample scattering power, the total exposure time, and the detector gain setting.

The Nickel dataset contains 11 images of powder diffraction rings with single crystal diffraction spots spreading throughout the patterns. The purpose of this dataset is to provide benchmark in accuracy and time performance of the machine learning algorithms used in this study. Since the images contain many single crystal diffraction spots, the dataset is ideal to be used to inspect different ML algorithms. In order to generate training and testing sets, the Auto Spot Mask (ASM) search of GSAS--II software is employed as the software is well-suited for single-crystal-diffraction-spot detection.

GSAS--II \cite{toby2013gsas} is a popular crystallography data analysis software application written in the Python programming language. Prior to determining the crystal structure of a material, GSAS--II enables users to remove irrelevant signals (e.g., single crystal diffraction spots) from 2D XRD images. This is facilitated by GSAS--II's ASM search capability. The ASM method checks one thin shell, expanded in the radial direction, at a time until it reaches the max 2-theta value used for integration, which is typically set close to the edge of the detector. Namely, the corners of the image are often ignored since it cannot make a circular thin shell without being trimmed. Once a thin shell is specified, the search will gather all of the pixel intensities within the thin shell and disregard the intensities that are larger than $F(\epsilon) = 100*erf(\epsilon/\sqrt{2})$, where $\epsilon$ is a user-defined hyperparameter with the range of 1--10. A smaller value of $\epsilon$ indicates more aggressive masking. Additionally, $erf$ is the error function. Standard deviation ($\sigma$) is calculated based on the intensities that are smaller or equal to $F(\epsilon)$. Subsequently, the ASM search will mask the pixels ($\boldsymbol{x}$) based on this condition:
\begin{equation} \label{GSASmask}
    \frac{(\boldsymbol{x}-median)}{\sigma} > \epsilon
\end{equation}
where the $median$ intensity is calculated without neglecting the intensities larger than $F(\epsilon)$.

Although the ASM search can detect the single crystal diffraction spots very well, it often fails to exclude preferred orientations or textures in the masking process. This shortcoming is because these characteristics also have high intensity, skewing the distribution of pixels within the thin shell further from the normal. In contrast to the spots, preferred orientations are desired signals in the 2D XRD pattern that should be taken into the integration process. Another disadvantage of using the ASM search is that the computational speed is relatively slow. It takes roughly 225 seconds to process one XRD image on a workstation with an Intel Xeon Silver 4110 2.10 GHz 32 core CPU. Accordingly, the expensive portion of the code can be re-written using the C Foreign Function Interface (CFFI) \cite{CFFI} Python library. By using the C programming language to process the highly time-consuming portion of the code, the computation time is reduced by 96\% to around 8.82 seconds per image on average. This improvement in speed may help to potentially enable XRD data processing on-the-fly during an experiment. Nevertheless, a speedup up to an order-of-magnitude can be achieved by ML methods as will be shown later.

The Battery-1 dataset contains 11 XRD images with single crystal diffraction spots and preferred orientations or textures. The dataset is certainly more complex than the Nickel dataset as multiple characteristics are present. Due to the charging/discharging nature of the experiment, distinct patterns can be traced in each of the images since each pattern may embody different signals from the charge carrier. Battery-2, Battery-3 and Battery-4 also are associated to the charging/discharging experimental conditions. Every dataset appertains to a different chemical composition and is collected under independent experimental conditions, ensuring variability in XRD patterns. Single crystal diffraction spots and textures are present in these datasets. For creating the training and testing sets, manual masking is carefully implemented to the single crystal diffraction spots only since the ASM search can falsely detect proper signals.

\subsection{Machine Learning Methods}

Support-vector machines (SVM) \cite{cortes1995support}, k-nearest neighbors (KNN) \cite{peterson2009k}, extra trees, random forest \cite{ho1995random}, and gradient boosting \cite{friedman2001greedy} algorithms are compared using 2D experimental XRD images. The SVM algorithm separates classes of labels by determining the decision boundaries, whereas the KNN technique is a voting algorithm that examines the n-closest neighbors to a data point and decides a class label based on its neighbors. The extra trees, random forest, and gradient boosting algorithms belong to the ensemble classification family. An ensemble algorithm typically generates many distinct models and averages the models to create more refined predictive results. In terms of model complexity, the ensemble algorithms are more complex than the SVM and KNN algorithms. Hence, they are expected to yield better predictive results in general.

Scikit-learn \cite{scikit-learn} --- a Python library for machine learning applications --- is utilized for the SVM, KNN, extra trees, and random forest algorithms. Multi-core CPU implementations are available in scikit-learn for all these methods except for of the SVM method. Since the Intel Xeon Silver 4110 2.10 GHz CPU used for this paper has 32 cores, 32 CPU cores will be employed for all training and testing procedures, apart from training and testing using the SVM method. The XGBoost library \cite{chen2015xgboost, Chen:2016:XST:2939672.2939785} is used for the gradient boosting framework. The XGBoost library allows the user to exploit multi-core CPUs as well as a Graphical Processing Unit (GPU). A GeForce RTX 2080 Ti GPU is used in this research. Results using both a multi-core CPU and GPU will be evaluated and compared in terms of time performance.

\begin{figure}
\centering
\caption{Plot illustrating the feature space of the 2D XRD pattern shown in Fig. \ref{xrd_example} in which every point belongs to a pixel.}
\includegraphics[width=0.95\textwidth]{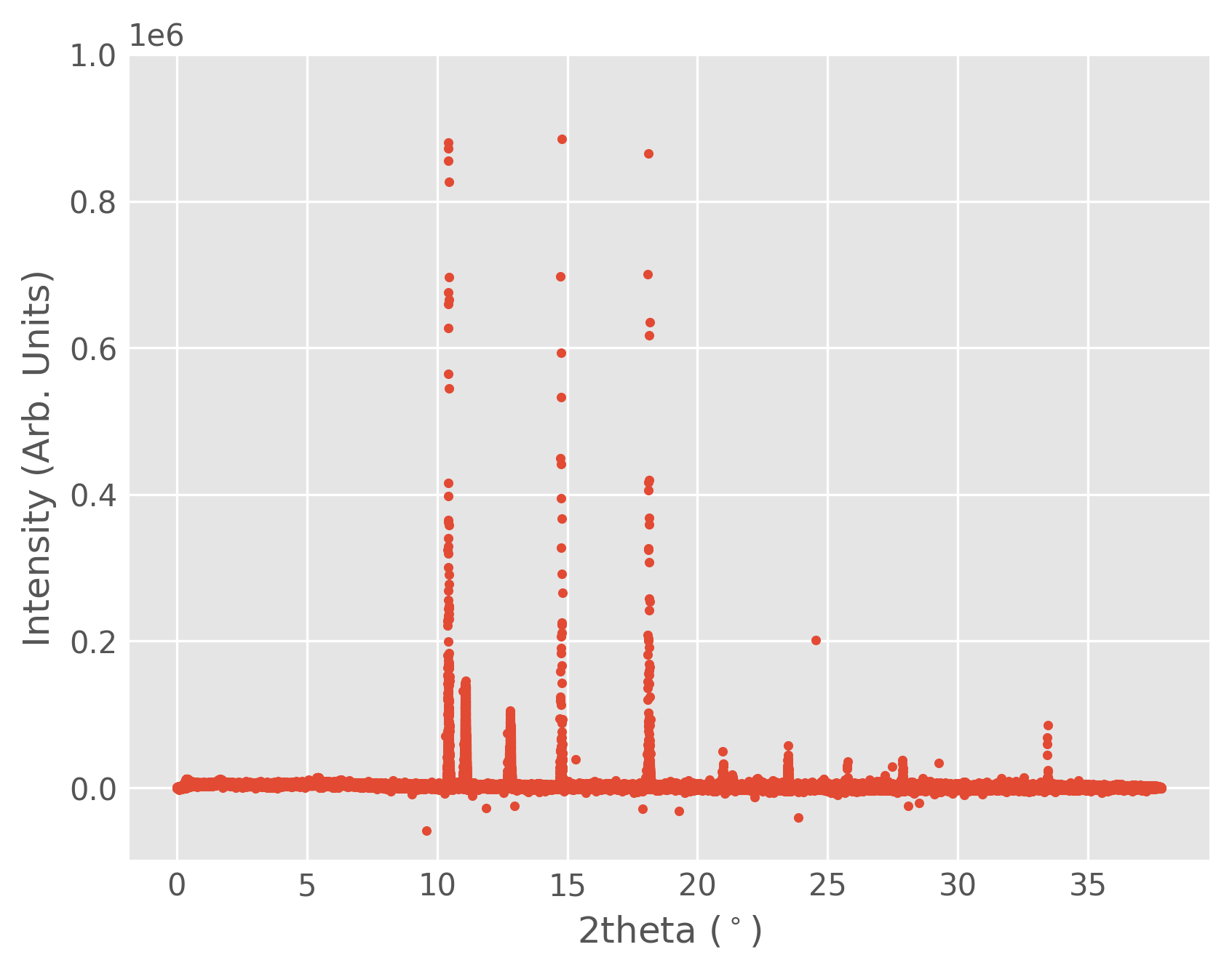}
\label{feature_space}
\end{figure}

All the ML methods require input values and return a class label as the output. The intensity and angle features are chosen as the input values while inspecting the ML methods with the Nickel dataset. Fig. \ref{feature_space} depicts the two feature spaces: intensity and angle. The angle information can be obtained by converting the metadata (e.g., wavelength, detector center, distance to detector, etc.) geometrically. There are 2880x2880 = 8,294,400 data points in one image. Consequently, the ML algorithms will also produce a 2D masking map with 8,294,400 outputs, yielding two label classes: 0 (no mask) and 1 (mask). Furthermore, a standard scaling normalization \cite{raju2020study} routine will be executed to all the XRD images in the training dataset prior to ML training to increase the robustness of the models, since the intensity values can be orders-of-magnitude higher than the angle values. One can avoid vanishing/exploding gradients in ML training if normalization schemata are appropriately implemented within a ML pipeline \cite{ioffe2015batch}.

The accuracy and time performances are evaluated among the 5 ML algorithms using the Nickel dataset. The comparison reveals auspicious algorithms that can handle 2D experimental XRD images. To attain fair accuracy, 5-fold cross validation will be applied within a grid search to find the optimal hyperparameters for each of the ML algorithms. Three random raw images are selected as a training dataset, and the remaining images are used as the test dataset for every fold. The averages of accuracy and time are applied within the 5-fold cross validation. The accuracy performance is measured by the recall and specificity scores:
\begin{equation} \label{recall}
    recall = \frac{true\:positive}{true\:positive + false\:negative}
\end{equation}
\begin{equation} \label{specificity}
    specificity = \frac{true\:negative}{true\:negative + false\:positive}
\end{equation}
The recall (True Positive or TP Rate) score can be considered as the ability of a machine learning classifier to find all the masked pixels. The specificity (True Negative or TN Rate) is shown as well, but not as an accuracy evaluation metric. When there is a label class imbalance (see Fig. \ref{xrd_example}b), the TP rate emphasizes the accuracy on the infrequent class, and the TN rate characterizes the classifier's performance over the larger negative class. Thus, a natural choice for accuracy is the recall or TP rate.

After juxtaposing the 5 ML algorithms, the trained gradient boosting model will be employed to predict the Battery-1 dataset. This assessment exposes the transferability of the trained model to a more complicated dataset, consisting of several characteristics. As the Battery-1 dataset has highly contrasting characteristics, it is expected that the trained model will perform poorly, since ML algorithms are known to perform well for interpolation but poorly for extrapolation. As shown in the Section \ref{result}, the trained model can be rectified by including a small subset of the Battery-1 dataset and the feature engineering approach to attain better predictive results. Feature engineering involves incorporating pixel locations which improves the performance of the gradient boosting model significantly. 

\begin{figure}
\centering
\caption{A schematic diagram of the ML pipeline for identifying single crystal diffraction spots.}
\includegraphics[width=0.95\textwidth]{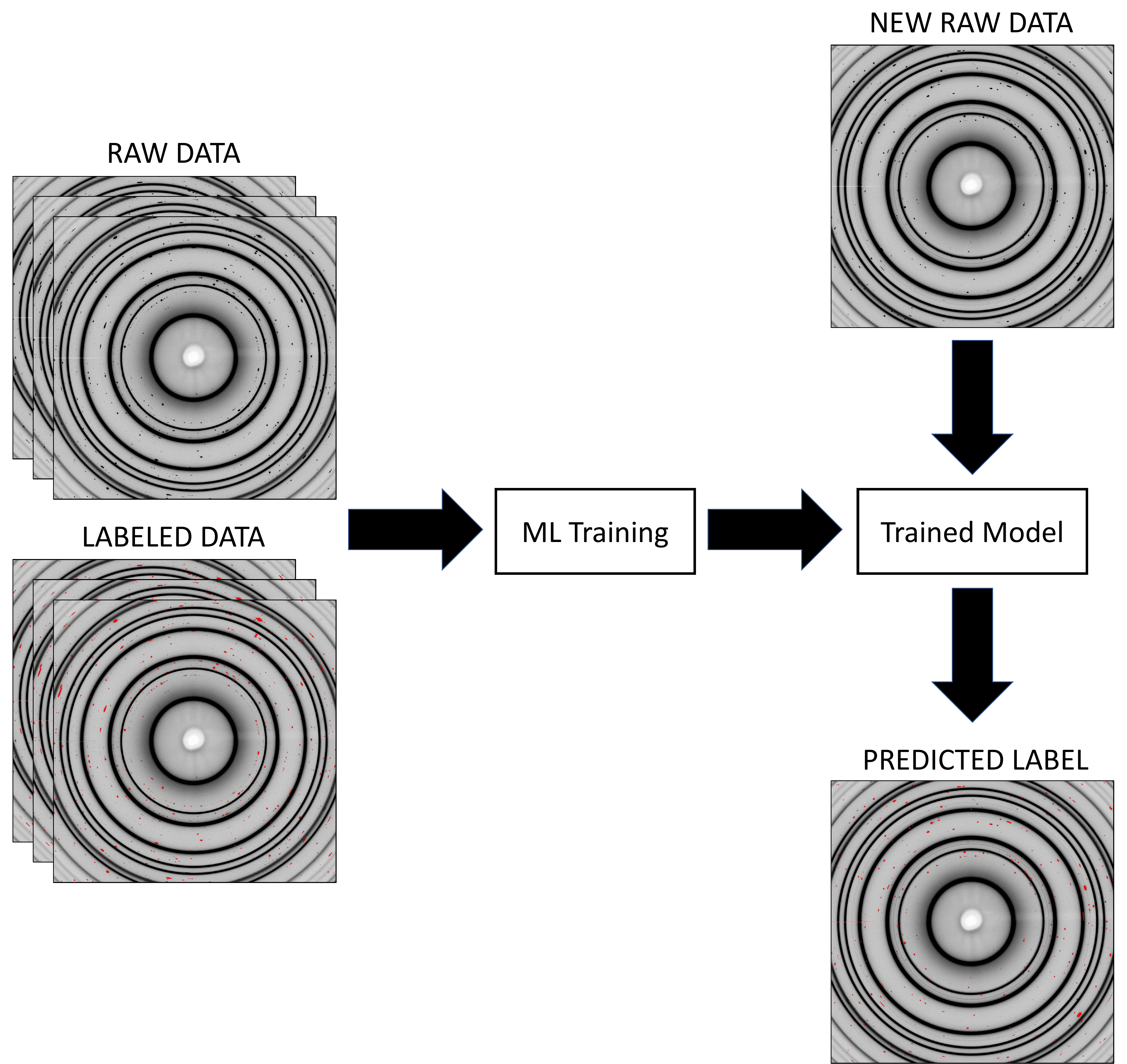}
\label{ml_xrd_workflow}
\end{figure}

Lastly, we will explore the predictive ability of the gradient boosting algorithm against all the datasets, providing diversity in chemical compositions, characteristics, and experimental conditions. In this case the Nickel, Battery-1, Battery-2, Battery-3, and Battery-4 datasets are used to assess the ML algorithms. Three images will be selected at random from each dataset as the training set, while the remaining images will be considered as the testing set, providing a wide diversity in the training and testing sets. The process will discover the model adaptability against various types of XRD images, belonging to different materials and collected under different experimental conditions. The schematic diagram of the ML pipeline used in this study is illustrated in Fig. \ref{ml_xrd_workflow}.

\section{Results and Discussion}\label{result}

\subsection{Machine Learning Algorithms Assessment}\label{ML_assessment}

Grid searches with 5-fold cross validations are applied to tune hyperparameters for the SVM, KNN, extra trees, random forest, and gradient boosting methods. The optimal hyperparameters are presented in Table \ref{optim_hyperparameters}. Most of the hyperparameter values lie on the aggressive side. However, there is no indication of overfitting according to the predicted results. Table \ref{nickel_table} summarizes the accuracy and time performances of the ML methods.

The SVM performances suggest that the method is not suited for this type of data. The method cannot find convergence given the training dataset. The training is set to terminate after 10,000 iterations. The algorithm fails to find boundary decision given the complex feature space (i.e., the feature space contains spikes of intensities). One can use a high-degree polynomial function as the kernel, instead of the Radial Basis Function. However, this can lead to overfitting. Moreover, the prediction time takes orders-of-magnitude longer than the other ML methods and the ASM search.

The KNN results are more encouraging. As mentioned above, the KNN algorithm predicts a new data point based on its neighboring data points. Consequently, it is preferable to choose an odd number of k-neighbors so that the algorithm can conclusively decide when there is an equal number of positive and negative votes. The TP and TN rates suggest that the KNN algorithm can produce reliable predictive results in comparison to the SVM method. The high TN rate shows that the method can sufficiently predict negatively labelled pixels. Meanwhile, the TP rate displays an adequate prediction, yielding 94.69\% accuracy despite the class imbalance. In spite of the reliable predictive capability, the time performance takes 4 times longer than the GSAS--II ASM search powered with the CFFI module. In this case, the predictive time shows that the algorithm is not ideal for on-the-fly masking in an XRD experiment. In addition, the predictive time of the algorithm is a function of the training dataset. This model's prediction time will increase as the size of the training dataset increases.

The ensemble algorithms show better predictive results in which the gradient boosting framework is the best of all the ML methods studied in this research. The extra tree method is able to capture the TP better than the KNN algorithm. However, the TN rate seems to perform slightly worse than the KNN algorithm. Nevertheless, the prediction time is greatly reduced by one order-of-magnitude, which is twice as fast as the GSAS--II ASM search with the CFFI module. The training time takes as long as the KNN algorithm. There is no distinct difference in TP rates between the extra tree and random forest methods. However, the random forest method takes longer to train on average, while the prediction time performance is the same for the two methods. 

The gradient boosting framework seems to perform the best in terms of the TP rate and time performance. It should be emphasized that the gradient boosting technique is the superior algorithm compared to the other methods in this study, so it is expected to outperform the others. The training time in a multi-core CPU and in a GPU are 2.23 seconds and 0.94 second, respectively. The time performances outclass the rest of the algorithms and the ASM search. Even with multi-core usage, the training time takes an order-of-magnitude less time than the other ML methods. If training time is not an issue, the gradient boosting framework can potentially be trained with many different types of raw XRD images collected from different experiments to further enhance the predictive power of the model. Furthermore, the gradient boosting method's (or the other ensemble methods) prediction time does not depend on the number of training points, so the prediction time remains the same as the training data size becomes larger. If a GPU is available, the gradient boosting framework can achieve the most efficient on-the-fly masking for an XRD experiment in comparison to the other algorithms in this study.

\subsection{Feature Assessment}\label{transferability}

We will inspect the transferability of the trained gradient boosting model that was previously trained as described in Subsection \ref{ML_assessment}. The trained model will be used to evaluate the Battery-1 dataset. Compared to the Nickel dataset, the Battery-1 dataset is more complex, as the images contain preferred orientations or textures, and single crystal diffraction spots. The GSAS--II ASM search is not able to distinguish among these characteristics. The previously trained gradient boosting model (i.e., only the Nickel dataset is included as a training dataset) also performs poorly in predicting the Battery-1 dataset. The poor prediction is expected as ML methods can reasonably interpolate data but poorly extrapolate data. Since the Battery-1 dataset was excluded in the previous gradient boosting training, the model predicts the overall TN and TP rates to be 97.3\% and 17.0\%.

\begin{figure}
\centering
\caption{Training time performance as a function of the number of training images based on the gradient boosting algorithm.}
\includegraphics[width=0.95\textwidth]{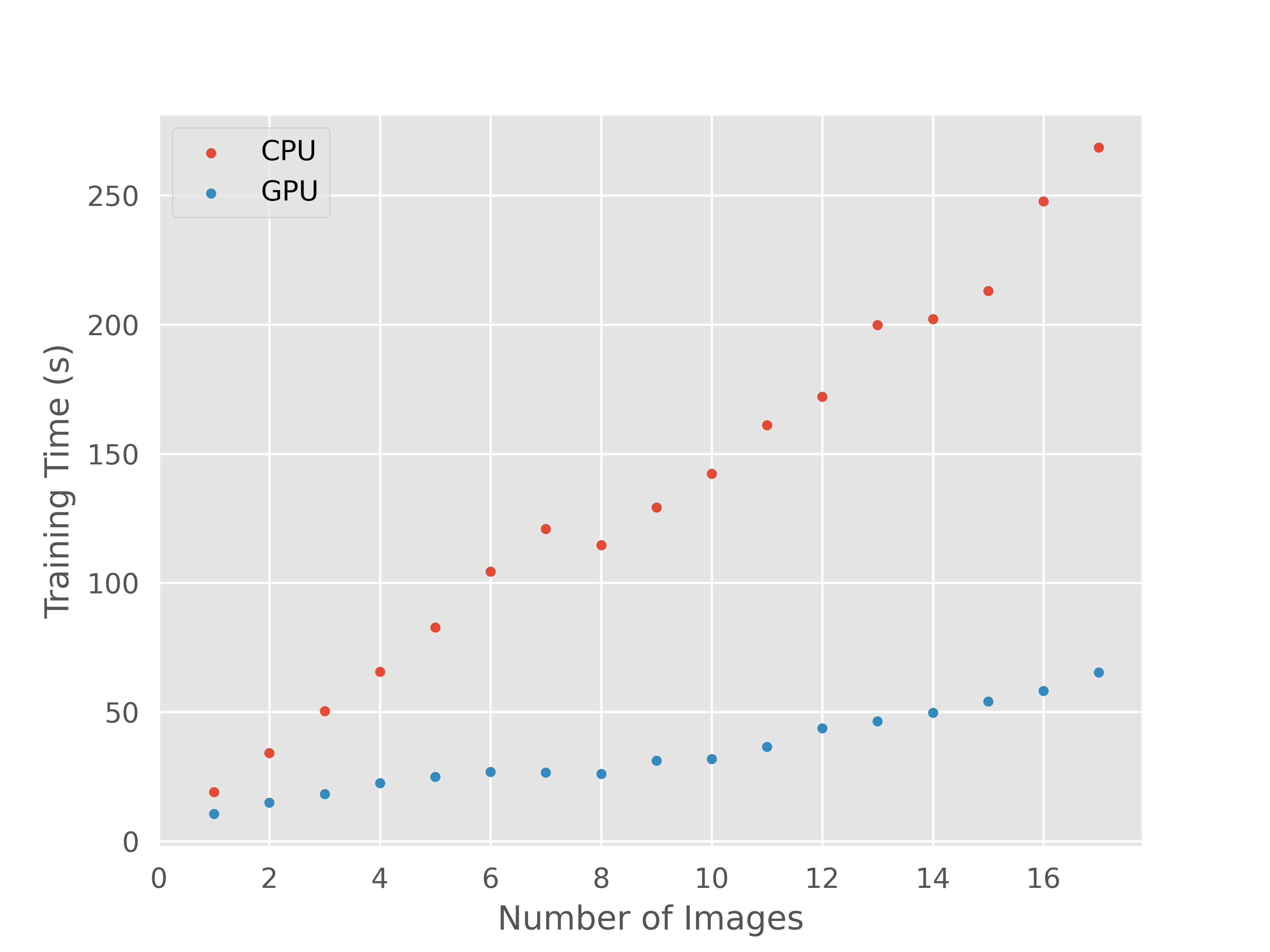}
\label{num_images}
\end{figure}

To increase the predicted TP rate of Battery-1 dataset, a small subset of the dataset is included into the training of the gradient boosting model. After including three images selected at random from the Battery-1 dataset along with the three random images from the Nickel dataset, the predicted TP rate of the Nickel dataset undergoes a modest reduction to 95.4\%, and the TN rate remains the same at 99.9\%. The training time with a multi-core CPU and with a GPU increases to 78.9 and 20.1 seconds, whereas the prediction time per image remains roughly unchanged at 2.31 seconds and 0.72 seconds for a multi-core CPU and for a GPU, respectively. Meanwhile, the overall TN and TP rates increase to 100.0\% and 54.1\%. More images of the Battery-1 dataset are included in the training as an attempt to increase the TP rate. At the maximum number of images, the TP rate increases to only 54.4\% as the training time increases linearly as a function of the number of images (see Fig. \ref{num_images}). The low TP rate of the Battery-1 dataset could be due to the fact that the images embody more complex characteristics that require a more powerful model to be able to generate finer results. 

In this case, feature engineering is used to strengthen the gradient boosting model by incorporating pixel locations as features. Again, three images are selected at random from each of the Nickel and Battery-1 datasets. This effort causes the predicted TP rate of Battery-1 dataset to increase to 97.6\%. As for the Nickel dataset, the TP rate also increases to 98.5\% from 95.4\%. The TN rates for both datasets remains at 100.0\%. For GPU usage, the training time slightly increases to 23.7 seconds, as compared to 20.1 seconds, previously, whereas the multi-core CPU training time increases to 89.4 seconds, as compared to 78.9 seconds, previously. The prediction time increases to 2.46 seconds and 1.07 seconds for a multi-core CPU and for a GPU, as compared to 2.31 seconds and 0.72 seconds, previously. Thereby, a small sacrifice in time performance can yield a significant improvement in the gradient boosting model through feature engineering by incorporating pixel locations as input features in the training.

\subsection{Diversity Assessment}\label{mixed_result_dataset}

\begin{figure}
\centering
\caption{TP rate results for different datasets. The solid curve represents GPU training time as a function of gradient boosting tree depth. The dashed curve represents multi-core CPU training time.}
\includegraphics[width=0.95\textwidth]{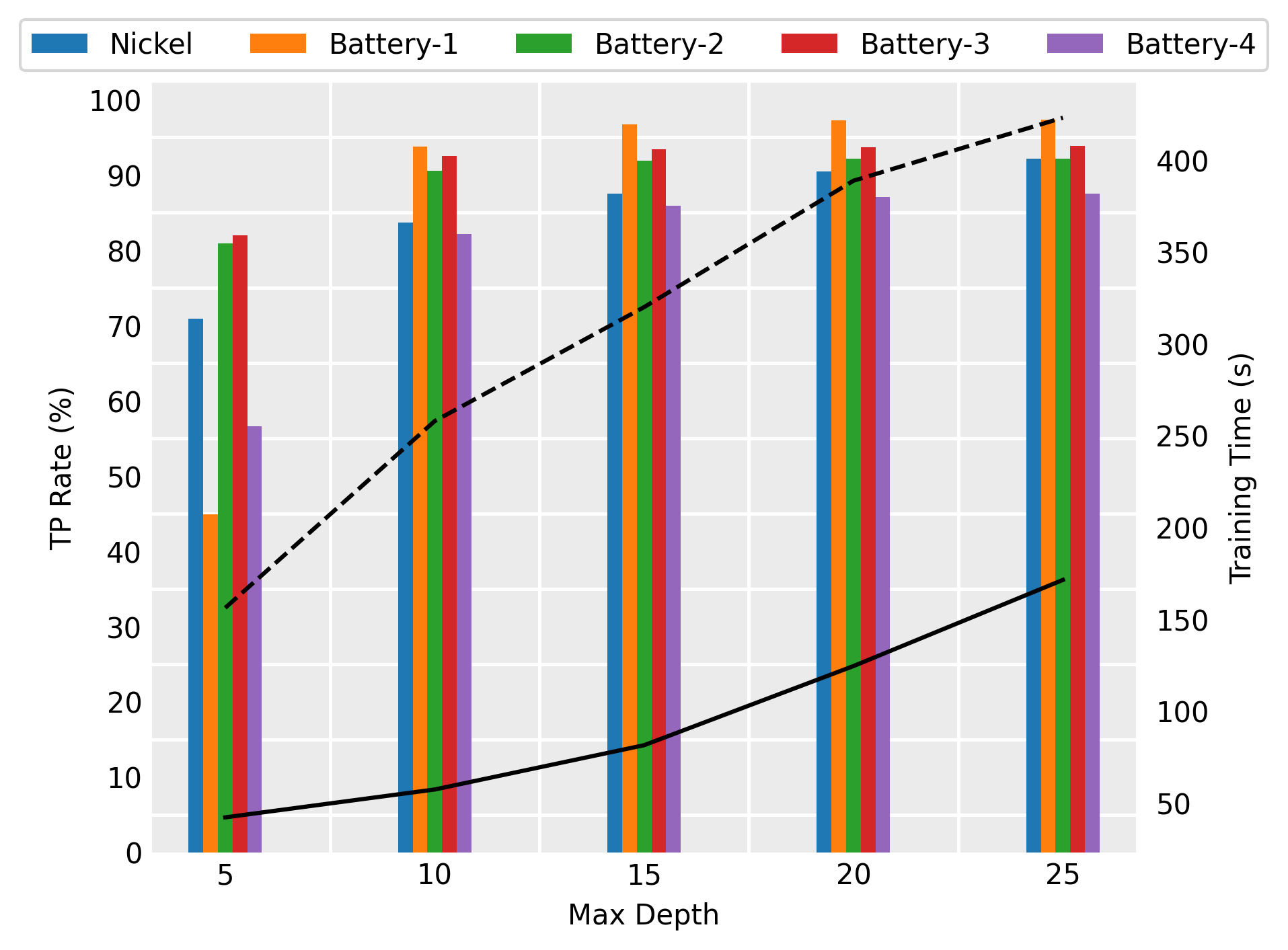}
\label{mixed_result}
\end{figure}

In this subsection, the gradient boosting model is employed to train various diverse datasets, Nickel, Battery-1, Battery-2, Battery-3, and Battery-4 datasets, collected under different experimental conditions as explained in Subsection \ref{datasets}. Fig. \ref{mixed_result} shows an increasing trend in TP rate as the gradient boosting method's maximum tree depth increases. Significant increases are detected when the maximum depth increases from 5 to 10. These effects are especially prominent for the Battery-1 and Battery-4 datasets due to both datasets having an order-of-magnitude fewer positive labels in comparison to the rest of the datasets, exacerbating the class imbalance issue. The sharp improvements as the maximum depth increases show that the model needs more flexibility (i.e., increasing the maximum depth of the model) to accurately describe all datasets. On the other hand, the TP rates show negligible improvements after a maximum depth of 20 given the training dataset, where the Nickel dataset exhibits the highest 1.7\% increase in TP rate. Overall, the gradient boosting model effectively predicts the Nickel, Battery-1, Battery-2, Battery-3, and Battery-4 datasets with high accuracy at 92.2\%, 97.4\%, 92.2\%, 93.9\%, and 87.5\%, respectively, at maximum depth of 25. The TN rates are almost 100\% on average for all datasets. The training time is shown in dashed and solid curves, representing the time measured with a multi-core CPU and with a GPU. Both training times increase as the maximum depth increases. However, CPU training time increases at slower rate while GPU training time indicates an exponential increase. The prediction time stays constant at 1.38 seconds on average with a GPU. On the other hand, the CPU prediction time increases as the maximum depth increases with the following trend: 1.85, 2.86, 3.63, 3.97, and 4.46 seconds.

The qualitative masking results can be found in Fig \ref{qual_visual}. It should be emphasized that the Nickel image only consists of two diffraction characteristics: single crystal diffraction spots, and powder rings without texture. Accordingly, the ASM search can identify the signals, and the gradient boosting method also yields reliable prediction. As for the rest of the images, the gradient boosting method identifies only the single crystal diffraction spots without recognizing the preferred orientations, while the ASM search fails to differentiate between the preferred orientations or textures and single crystal diffraction spots. The radii of the single-crystal-diffraction-spot masks generated by the gradient boosting method are more tightly constrained than those generated by the ASM search approach. The phenomena can be clearly observed in the Battery-2 and Battery-3 images that the gradient boosting masks have smaller radii compared to ASM masks, in general. The circumstance can be caused by the skewed datasets, in which the images contain more negative labels than positive labels. Nevertheless, all of the single-crystal-diffraction-spot centers are found by the gradient boosting method.

\section{Conclusion and Future Work}

In this work, the effectiveness and efficiency of ML methods, such as SVM, KNN, extra trees, random forest, and gradient boosting algorithms, to identify and separate single crystal diffraction spots in 2D powder XRD images---containing more than 8 million pixels---to enable precise analyses of the 1D powder diffraction patterns, is investigated. It is shown that KNN, extra trees, random forest, and gradient boosting display accurate predictions yielding more than 95\% of TP rate and 99.9\% of TN rate for the Nickel dataset. The gradient boosting method produces the best time and accuracy performance while offering both multi-core CPU and GPU support. The gradient boosting method has shown that it can reliably predict the images that are not present during the training, yielding TP rates in the range of 87.5\% to 97.4\% when it was tested with all of the datasets. In addition, the TN rates remain at nearly 100\% on average. The training time increases linearly as a function of training data size on both a multi-core CPU and on a GPU, where the GPU training time increases at slower pace than the multi-core CPU training time. It is also observed that the GPU prediction time of the gradient boosting method remains constant at 1.38 seconds despite the increase in model complexity (i.e., increasing maximum depth) and the size of the training dataset. The CPU prediction time experiences a slight increase as the model complexity increases, and it remains the same as training data size increases.

\begin{figure}
\centering
\caption{Qualitative masking results. (a)-(e) are the raw images. (f)-(j) are the results of masking using ASM search. (k)-(o) are results of masking using the gradient boosting method. Each row consists of images that belong to Nickel, Battery-1, Battery-2, Battery-3 or Battery-4 dataset, respectively from top to bottom.}
\includegraphics[width=0.95\textwidth]{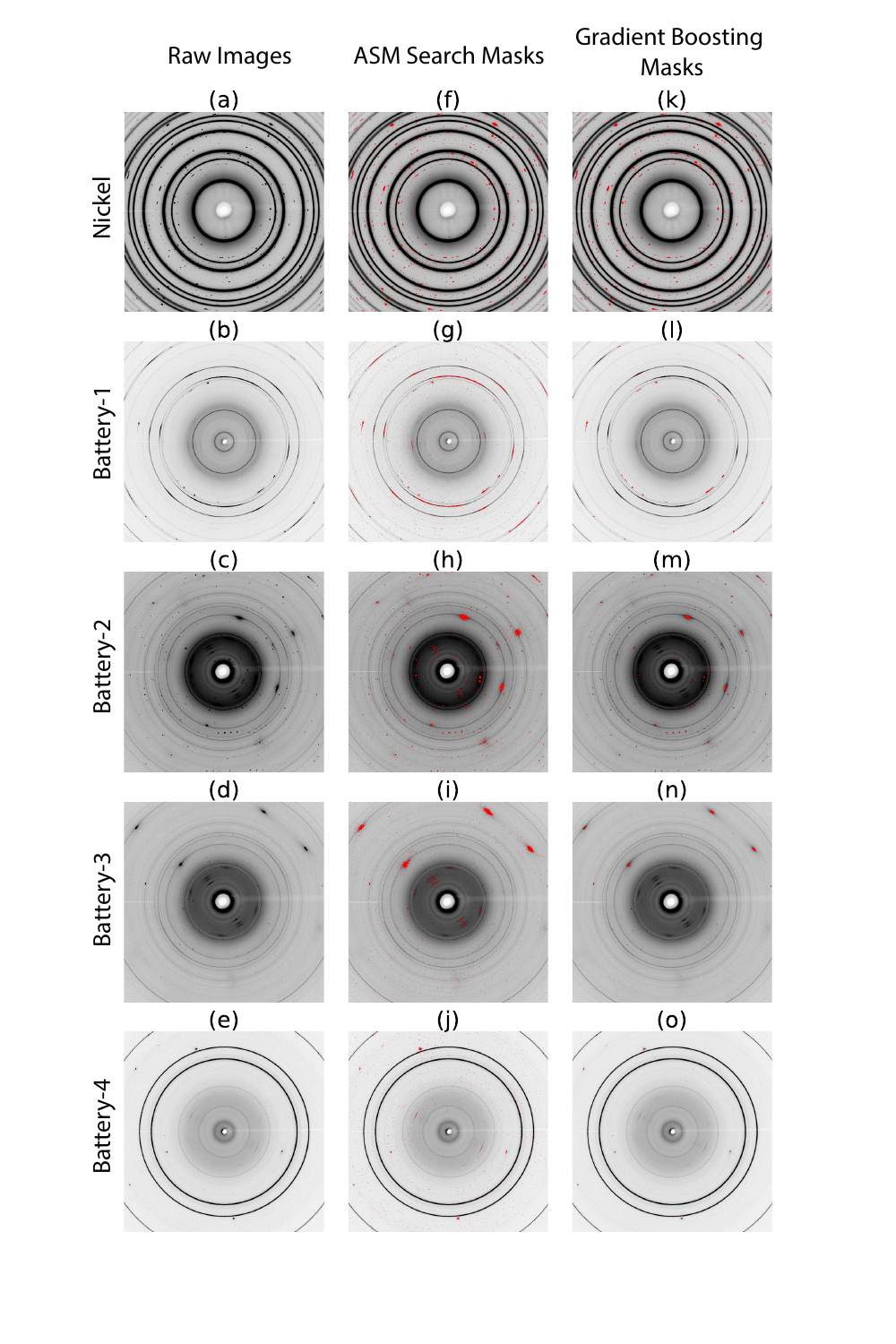}
\label{qual_visual}
\end{figure}

There are limitations to using the gradient boosting method for the identification and separation of single crystal diffraction spots in XRD images. First, the XRD datasets are skewed representing class imbalance, in which the datasets have orders-of-magnitude more negative labels than the positive labels. As it was found in Fig. \ref{qual_visual}, the masks of the gradient boosting predictions exhibit smaller masking radii at the single crystal diffraction spots, yet it can better differentiate among artifacts and other characteristics in comparison to the ASM search approach. Second, extrapolation is a known drawback of ML methods. In this case, the gradient boosting method will fail to identify the single crystal diffraction spots in XRD images where the images have substantially stronger pixel intensities. A small subset of XRD images needs to be included in the model training if it has not been incorporated in the previous training as shown in Subsection \ref{mixed_result_dataset}.

We plan further investigations to enhance this ML-based artifact identification approach. The standard scaling normalization used in this paper confines the intensities to those found in the training dataset. We plan to explore other normalization techniques that do not bound the intensities to the training dataset. Additional feature engineering approaches, such as using edge detection will be explored. Further, we plan to study the application of CNNs to increase TP rate and transferability.

The application of ML methods, particularly the gradient boosting method, offers major improvements in both the effectiveness and efficiency relative to conventional and manual artifact identification methods. These ML methods offer substantial potential advancements in addressing the bottleneck in XRD artifact identification and reduce human intervention required during an experiment, opening the possibility of on-the-fly data processing capabilities during an experiment. The code used in this study is available at https://github.com/AdvancedPhotonSource/airxd-ml.



\ack{Acknowledgements}

The authors thank Dr. Olaf J. Borkiewicz, Dr. Andrey A. Yakovenko, and Dr. Tiffany L. Kinnibrugh for their valuable insights at the initiation of the project, and Dr. Alexander Hexemer and Dr. Subramanian Sankaranarayanan for their full support.

This work is supported by the U.S. Department of Energy (DOE) Office of Science-Basic Energy Sciences award Collaborative Machine Learning Platform for Scientific Discovery. This research used resources of the Advanced Photon Source, a U.S. DOE Office of Science user facility at Argonne National Laboratory and is based on research supported by the U.S. DOE Office of Science-Basic Energy Sciences, under Contract No. DE-AC02-06CH11357.

\referencelist{}

\begin{table}\label{datasets_table}
\centering
\caption{Characteristics of datasets used for ML algorithm comparisons. SCD spots and POs stand for single crystal diffraction spots and preferred orientations, respectively.}
\begin{tabular}{lccc}
\hline\hline
 Name           & No. of Images & Characteristics               & Experimental Condition\\
\hline
 Nickel         & 11            & SCD spots                     & Temperature ramping\\
 Battery-1      & 11            & SCD spots and POs             & Charging/discharging\\
 Battery-2      & 12            & SCD spots and textures        & Charging/discharging\\
 Battery-3      & 12            & SCD spots and textures        & Charging/discharging\\
 Battery-4      & 12            & SCD spots and textures        & Charging/discharging\\
\hline\hline
\end{tabular}
\end{table}

\begin{table}\label{optim_hyperparameters}
\centering
\caption{The optimal hyperparameters of the SVM, KNN, extra trees, random forest, and gradient boosting methods. Default values are used if they are not listed.}
\begin{tabular}{llc}
\hline\hline
 & Hyperparameters        & Value   \\
\hline
 KNN            & n\_neighbors          & 3         \\
                & algorithm             & auto      \\
                & leaf\_size            & 5         \\
                & p                     & 2         \\
\hline
Extra Trees     & n\_estimators         & 35        \\
                & criterion             & gini      \\
                
\hline
Random Forest   & n\_estimators         & 35        \\
                & criterion             & gini      \\
                & min\_samples\_split   & 2         \\

\hline
Gradient Boosting   & n\_estimators         & 35        \\
                    & max\_bin              & 10000     \\
                    & min\_child\_weight    & 1         \\
                    & max\_depth            & 10        \\

\hline\hline
\end{tabular}
\end{table}

\begin{table}\label{nickel_table}
\centering
\caption{Benchmarking results using the Nickel dataset. TN and TP rates represent the true negative and true positive rates evaluated with the Nickel test set. The training/testing time performances are measured with CPU(GPU).}
\begin{tabular}{lccc}
\hline\hline
 Algorithms         & TN Rate(\%)    & TP Rate(\%)    & Time(s)\\
\hline
 SVM                & 78.45$\pm$0.23             & 46.17$\pm$3.31             &2360/1450\\
 KNN                & 99.93$\pm$0.13             & 94.69$\pm$1.01             &218/41.6\\
 Extra Trees        & 99.88$\pm$0.17             & 97.73$\pm$0.68             &255/4.83\\
 Random Forest      & 99.96$\pm$0.06             & 97.33$\pm$0.30             &503/4.53\\
 Gradient Boosting  & 99.94$\pm$0.10             & 98.22$\pm$0.39             &42.5(11.3)/2.23(0.94)\\
\hline\hline
\end{tabular}
\end{table}

\end{document}